\documentclass[12pt]{article}
\newcommand{\rot}{rot}
\renewcommand{\d}{\displaystyle}
\begin{document}

\begin{center}
{\bf V. S.~Yarunin}\\[0.5cm]
{\bf Joint Institute for Nuclear Research, Moscow Region,
141980, Dubna, Russia }\\
{\bf E-mail: yarunin@thsun1.jinr.ru}\\[3mm]
{\large\bf ODD-ODD MAGNETIC INTERACTION\\
AND SPONTANEOUS ORTHO-PARA TRANSITIONS \\
IN MOLECULE AND MOLECULE HYDROGEN ION}\\[3mm]
{\bf Leningrad University Bulletin n 4\\
Series of Physics and Chemistry, Issue 1, p.16-22\\
Leningrad, 1966 (in Russian)}
\footnote{Translated into English by V.S.~Yarunin, 2001.}\\[2mm]

1. INTRODUCTION\\

\end{center}

Molecular $H_2$ is a mixture of ortho- and para- modifications,
a sum of nuclear spins in which are 1 and 0 correspondingly. The equilibrium
concentration of these modifications depends on temperature.
A mixture at 273 $K$ contains  25\% of para-hydrogen, while close to 0 $K$
a mixture contains 100\% of para-hydrogen. It was found, that a change
of temperature leads to a thermal equilibrium during a long time of
relaxation, but this time may be diminished by the use of catalyst.
Then, an ortho-para transition in $H_2$ is possible due to paramagnetic
admixture and thermal dissociation and recombination of $H_2$ molecules
at large temperatures 800--1000 $K$ [1]. The review on the orto-
para-hydrogen investigations may be found in [2].

A radiation via the ortho-para transitions is impossible, as a rotating
and spin states of nuclei in molecule has a certain permutation symmetries
of nuclear space and spin coordinates. Really, the wave function of $H_2$
is anti-symmetrical relative the exchange the places of atoms.
Let us suppose, that the nuclear spin, rotative and oscillative states
are separated as
\begin{equation}
\psi_{mol}=\psi_{el}(r)\psi_{el}(\sigma)\psi_{osc}(R_1,R_2)
\psi_{rot}(R_1,R_2)\psi_{nuc}(\Sigma_1,\Sigma_2),
\end{equation}
here $r$ and $\sigma$ are the coordinates and spins of electrons,
$R_1,R_2$ and $\Sigma_1,\Sigma_2$ are the coordinates and spins of nuclei.
In this case a para-$H_2$ state is described
by an antisymmetrical
nuclear $\psi_{nuc}(\Sigma_1,\Sigma_2)$ wave function and by a symmetrical
rotational $\psi_{rot}(R_1,R_2)$, while ortho-state is described
by a symmetrical nuclear $\psi_{nuc}(\Sigma_1,\Sigma_2)$ wave function
and by an anti-symmetrical rotational $\psi_{rot}(R_1,R_2)$.

Let us consider an electric dipole $P$ matrix elements over the nuclear
wave functions
$$
<\psi_{nuc}^+|P|\psi_{nuc}^->,\quad <\psi_{nuc}^-|P|\psi_{nuc}^+>,
$$
where $\psi^+$ and $\psi^-$ are the symmetrical and antisymmetrical
states relative a nuclei permutation.

These matrix elements are equal to $0$, as they must be invariant
accordingly nuclear permutation in molecule and
they must change their signers at the same time. Therefore, the ortho-para
transitions are impossible between the states (1). This rule is valid
for all the two-atomic (with the same atoms) molecules with non-zero spin.

An effort to measure an experimental possibility of a spontaneous
ortho-para transition in $H_2$ was done in [1]. A time for a half-number
transition (if exists) was found to be more 1 year.

There is an estimation by Wigner of the radiative ortho-para transition
probability in $H_2$, unpublished by him but quoted in [1]. Wigner's
idea was that an nuclear spin interaction breaks the symmetry of
unperturbed states (1). The probability of such a transition was
found by Wigner as one per 300 years with as assumption that
nuclear perturbation in $H_2$ may be estimated as $1000^{-1}$ of $^4$He
multiplet.

Orto-para transitions in solid $H_2$ via a magnetic interaction between
ortho-molecules were considered in [3].

In this paper the interactions are considered,
providing spontaneous radiative transitions for $H_2$ and, in more
details, for $H_2^+$.

\begin{center}
2. NUCLEAR MAGNETIC INTERACTION AND ORTHO-PARA\\
TRANSITIONS IN $H_2$
\end{center}

Radiative ortho-para transitions are impossible for $H_2$ in an
approximation (1), but there are
interactions in $H_2$, that are not taken into account by (1).
In order to look for interaction, appropriate for a broken symmetry
of (1), let us take the whole Schr\"{o}dinger equation for $H_2$
\begin{equation}
(K+V_{el}+V_{nuc-nuc}+V_{el-nuc})\psi=E\psi,
\end{equation}
where $K$ is a kinetic energy of electrons and nuclei, $V_{el}$ is
electron interaction, $V_{nuk-nuk}$ is nuclear interaction and $V_{el-nuk}$
is the electron-nuclear interaction.

If operator $V=V_{el}+V_{nuc-nuc}+V_{el-nuc}$ contains only electrostatic
interactions $(V=V(r,R))$, molecular wave function may be taken as (1),
so as nuclear motion is slow comparatively electronic and rotational motion
is slow comparatively oscillative.

Let us consider magnetic interactions. In this case $V(r,\sigma,R,\Sigma)$
is symmetric relative the nuclear permutation, that means the
simultaneous permutation both space and spin nuclear coordinates.
Still the odd-odd part $V_{aa}$ may be selected in $V$, that is
antisymmetric over permutation of nuclear space or spin coordinates
$$
V=V_{aa} +V_{ss},
$$
$$
V_{aa}=\frac{1}{4}\left[V(R_1 R_2 \Sigma_1 \Sigma_2)
-V(R_2 R_1 \Sigma_1 \Sigma_2)-V(R_1 R_2 \Sigma_2 \Sigma_1)+
V(R_2 R_1 \Sigma_2 \Sigma_1)\right].
$$
An interaction $V_{aa}$ changes a symmetry of nuclear space and nuclear
spin states separately. This change gives an opportunity to consider $V_{aa}$
as a perturbation, that makes an optical
ortho-para transition in $H_2$ possible.

Unperturbed states are taken as the eigenfunctions of (2) without magnetic
fields in approximation (1)
\begin{equation}
\psi_{para}=\psi_{el}(r)\psi_{el}(\sigma)\psi_{osc}(R_1,R_2)
\psi_{rot}^+(R_1,R_2)\psi_{nuc}^-(\Sigma_1,\Sigma_2)=\psi_{pvr}^-,
\end{equation}
\begin{equation}
\psi_{ortho}=\psi_{el}(r)\psi_{el}(\sigma)\psi_{osc}(R_1,R_2)
\psi_{rot}^-(R_1,R_2)\psi_{nuc}^+(\Sigma_1,\Sigma_2)=\psi_{pvr}^+.
\end{equation}

Quantum numbers $p$ are electronic, $v$ -- oscillative, $r$ -- rotative.
Operator $V_{ss}$ perturbs the states (3,4) without changing the
symmetry of nuclear, spin and spatial states. It means, that
$V_{ss}$ doesn't provide the possibility of ortho--para transitions.
$$
<\psi^{\pm}|V|\psi^->=<\psi^-|V_{ss}|\psi^-> + <\psi_+|V_{aa}|\psi^->.
$$
The perturbed wave function $\Phi$ of the ground (para) state of $H_2$
looks like
$$
\Phi_{para}=\psi^-_{000} + \phi^+,
$$
\begin{equation}
\phi^+=\sum_{pvr}\frac{<\psi_{pvr}^+|V_{aa}|\psi_{000}^->}
{E_{000}^-E_{pvr}^+} \psi_{pvr}^+.
\end{equation}
Ground state wave function $\Psi_{000}^{-}$ is not degenerated.
On finding  the perturbed wave functions $\Phi$ of excited states
of molecule one should take into account the degeneration of the
wave functions (3, 4) on spin of electrons, on projection of angular
momentum of electrons on molecule axis and, for the ortho-states,
on spin of nucleus.

The excited state wave function of molecule
$\Phi (\vec{r}, \vec{\sigma}, \vec{R}, \vec{\Sigma})$ is antisymmetric
relative the nuclei permutation. The difference between functions $\Phi$
and $\Psi$  consists in the fact, that
$\Psi$ has the term of opposite symmetry on spatial and spin
coordinates of nuclei. Therefore, the matrix elements
$$
\int \Phi_{para}^{*}P\Phi_{ortho}d\tau,
\quad
\int \Phi_{ortho}^{*}P\Phi_{para}d\tau,
$$
are not equal to zero due to the presence of terms of the same symmetry
on spins of nuclei and  coordinates  permutation   in $\Phi_{ortho}$
and $\Phi_{para}$. Thus, the odd--odd part of magnetic interaction
 operator in molecule excites the symmetry of pure ortho-- and para--states
 and ortho--para transitions becomes possible.
 The conclusion on ortho--para transitions prohibition breaking
 due to the interaction $V_{aa}$ caused by non-additive dependence $V_{aa}$
 on spatial coordinates and spins of nuclei.

\begin{center}
{\bf 3. EXCITATION OF ORTHO-- AND PARA--STATES IN MOLECULAR ION OF
HYDROGEN.}
\end{center}

The conclusion of the last paragraph on ortho--para transitions
prohibition breaking due to odd--odd part of magnetic interaction in
molecule remains true for molecular ion $H^+_{2}$. On the other hand
the example of hydrogen's molecular ion allow us to follow up
the role of different magnetic interactions on calculating
a perturbed wave functions.

The matrix elements for perturbed  ground para--state of molecular ion
$\Phi_{para}$ take the form
$$
<\psi^{+}_{pvr}|V_{aa}|\psi^{-}_{000}>.
$$
As against the case of molecule $H_2$, the unperturbed states of
$H^+_2$ have double degeneracy on spin of electron $s=\frac{1}{2}$.
Formula (5) will be true if we put the electron spin projection
into the set of quantum numbers.

The direct interaction between magnetic momenta of nuclei in $V_{n-n}$
and magnetic interaction between electron and nuclei in $V_{e-n}$
give a contribution to the operator $V_{aa}$. The first one
is proportional to $m_p^{-2}\,$ ($m_p$ is the mass of proton).
The second one is proportional to $m_p^{-1}$. It will be seen later,
that in the case of $H^{+}_2$ ion  the contribution of the first term
is small and may be neglected on calculating the perturbed wave
functions.

The energy operator of electron--electron magnetic interaction
can be found from Dirac's equation for electron in Pauli form [4]
$$
{(V_{e-n})}_{magn}=\frac{ec}{E+E_{0} +e\varphi}\left[ h\left(
\vec{{\bf H}}\,\vec{\sigma}\right)+\left(\vec{p }\,\vec{{\bf A}}\right)+
\left(\vec{{\bf  A}}\,\vec{ p}\right)\right]+
$$
\begin{equation}
\frac{e^2 hc(-i)}
{{(E+E_0+e\varphi)}^2}\left(\vec{\sigma}\,\vec{{\bf E}}\right)
\left(\vec{\sigma}\,\vec{{\bf A}}\right) .
\end{equation}
The terms proportional to ${\bf A^2}$ are neglected here;
$E+E_0=2mc^2+\omega,\quad \omega$ is nonrelativistic part of electron
energy; $\vec{\sigma}$ is the electron spin operator; $\vec{p}$ is
the electron momentum operator.

If we consider, that the velocity of nucleus is much smaller then
the velocity of electron, we can neglect the influence of molecular
ion rotation on electron movement. Then, the field generated by protons
in coordinate set connected with rotating nuclei can be written in the
form
\begin{equation}
\left.
\begin{array}{l}
{\displaystyle \vec{{\bf A}}=\frac{[\vec{\mu}_1\,\vec{r}_1]}{r_1^3}+
\frac{[\vec{\mu}_2\,\vec{ r}_2]}{r_2^3}, \quad \varphi=\frac{e}{r_1}+
\frac{e}{r_2}},\\
\phantom{dddddd}\\
{\d\vec{{\bf E}}=\frac{e\vec{r}_1}{r_1^3}+
\frac{e\vec{ r}_2}{r_2^3},}\\
\phantom{dddddd}\\
{\d\vec{{\bf H}}=\rot\vec{{\bf A}}=\frac{3\vec{r}_1
(\vec{\mu}_1\,\vec{ r}_1)}{r_1^5}-\frac{\vec{\mu}_1}{r_1^3}
+\frac{3\vec{r}_2
(\vec{\mu}_2\,\vec{ r}_2)}{r_2^5}-\frac{\vec{\mu}_2}{r_2^3},}
\end{array}
\right\}
\end{equation}
$\vec{r}_1=\vec{r}-\vec{R}_1,\quad \vec{r}_2=\vec{r}-\vec{R}_2,\, \vec{r}$
is the coordinate of electron; $\vec{R}_1, \vec{R}_2$ are the
coordinates of nuclei; $\vec{\mu}_1, \vec{\mu}_2$ are the magnetic momenta
of nuclei. The coordinate set is placed in a half of the distance
between nuclei.

Putting (7) into (6) one can write the part $W$ of magnetic
interaction operator between electron and protons ${(V_{e-n})}_{magn}$,
which gives the contribution to $V_{aa}$ in the form,
$$
W=a\left[ \frac{\left(\vec{\Sigma}_1[\vec{r}_1\,\vec{p}\,]\right)}{r_1^3}+
\frac{\left(\vec{\Sigma}_2[\vec{r}_2\,\vec{p}\,]\right)}{r_2^3}+
\frac{h}{2}\left(
3\frac{\left(\vec{\Sigma}_1\,\vec{r}_1\right)\left(\vec{r}_1\,
\vec{\sigma}\right)}{r_1^5}-
\frac{\left(\vec{\Sigma}_1\,\vec{\sigma}\right)}{r_1^3}+
\right.\right.
$$
$$
\left.\left.
+3\frac{\left(\vec{\Sigma}_2\,\vec{r}_2\right)
\left(\vec{r}_2\,\vec{\sigma}\right)}{r_2^5}-
\frac{\left(\vec{\Sigma}_2\,\vec{\sigma}\right)}{r_2^3}\right)\right]+
b\left[
\frac{\left(\vec{\Sigma}_1\,\vec{\sigma}\right)}{r_1^4}-
\frac{\left(\vec{r}_1\,\vec{\Sigma}_1\right)
\left(\vec{r}_1\,\vec{\sigma}\right)}{r_1^6}+
\frac{\left(\vec{\Sigma}_2\,\vec{\sigma}\right)}{r_2^4}-
\right.
$$
$$
\left.
-3\frac{\left(\vec{\sigma}\,\vec{r}_2\right)
\left(\vec{r}_2\,\vec{\Sigma}_2\right)}{r_2^6}-
\frac{1}{r_1^3 r_2^3}\left\{
\left(\vec{\sigma}\,\vec{r}_1\right)
\left(\vec{r}_2\,\vec{\Sigma}_1\right)+
\left(\vec{\sigma}\,\vec{r}_2\right)
\left(\vec{\Sigma}_2\,\vec{r}_1\right)\right\}+
\right.
$$
\begin{equation}
\left.
+\frac{i}{r_1^3 r_2^3}
\left(\vec{\Sigma}_1\,\vec{\Sigma}_2\right)[\vec{r}_2\,\vec{r}_1\,]\right],
\end{equation}
$$
a=\frac{2ec\mu_p}{E+E_0+e\varphi},\quad
b=\frac{e^3 hc\mu_p}{{\left(E+E_0+e\varphi\right)}^2},\quad
\vec{\mu}_1=\mu_p\,\vec{\Sigma}_1,\quad
\vec{\mu}_2=\mu_p\,\vec{\Sigma}_2.
$$
The terms placed inside the square brackets describe the
interactions between spins of nuclei and spin of electron
and between spins of nuclei and angular momentum of electron.
To the present moment we consider the matrix elements for perturbed
wave functions of ortho-- and para--states without taking into account
the influence of molecular ion rotation on ortho-- and para--transitions.
In particular, the approximation (1) for unperturbed wave function is
an adiabatic approximation for the spatial wave function of electron
$\psi_{el}(\vec{r})$. The coordinates of nuclei $\vec{R_1},\, \vec{R_2}$
are contained in $\psi_{el}(\vec{r})$ as a parameters.
In order to calculate the matrix elements over rotated  states
one should integrate it over the parameters $\vec{R_1},\, \vec{R_2}$.
The influence of nuclei rotation on electron motion and other rotated
interactions can be taken into account by putting the corresponding
terms into  complete expression for ion interaction operator $V$.

It turns out, that nonadiabatic term in ion hamiltonian
$\int \psi_{el}\Delta_{n}\psi_{el}$, which prevents the separation
of electron--nuclei motion doesn't depend on spins of nuclei
$\vec{\Sigma_1}, \vec{\Sigma_2}$ and therefore gives the contribution
to $V_{ss}$ but not to $V_{aa}$. So, the adiabatic approximation
doesn't influence on the possibility of ortho--para transitions.

Having interested in interactions which influence on the possibility
of ortho--para transitions and are proportional to $m_p^{-1}$, one
can calculate magnetic interaction term in ion $H_2^{+}$ taking into
account the rotations. For this purpose, one should define the potentials
created by protons rotation in immovable coordinate set and put it
into (6) adding to (6) a term from  $-e\varphi$.
One can use the formulas for electromagnetic field transformation
on changing a coordinate set.
$$
\left\{
\begin{array}{l}
{\d \vec{{\bf E}}\,'=\vec{{\bf E}}+\frac{1}{c}\left[
\vec{{\bf H}}\,\vec{v}\,'\right]},\\
\phantom{dddddd}\\
{\d \vec{{\bf H}}\,'=\vec{{\bf H}}-\frac{1}{c}\left[
\vec{{\bf E}}\,\vec{v}\,'\right]},
\end{array}
\right.\quad
\left\{
\begin{array}{l}
{\d \vec{{\bf A}}'=\vec{{\bf A}}+\frac{1}{c}\vec{ v}\,' \varphi},\\
\phantom{ffffff}\\
{\d\varphi' =\varphi +\frac{1}{c}\left(\vec{v}\,'\,\vec{\bf A}\right)},
\end{array}
\right.
\quad
\vec{v}\,'=\frac{\left[\vec{J}\,\vec{r}\right]}{I},
$$
$I$ is the mechanical moment of molecular ion. Touched variables denote
a field in immoveable coordinate set and variables
$\vec{{\bf H}},\,\vec{{\bf E}},\,\vec{{\bf A}},\, \varphi$ describe
a field in a rotated one connected with the nuclei and have the form (7).
$\vec{J}$ is the operator for rotated mechanical moment of nuclei.
The last one commutates with operators without molecular ion
rotation variables.

The interaction between electron and protons has the form
$$
{(V_{e-n})}_{magn}^{'}=-\frac{e}{c}\left( \vec{v}_1\,'\,\vec{{\bf A}}\right)
+\frac{ec}{E+E_0 +e\varphi}\left[
\left(\vec{p}\,'\,\vec{{\bf A}}'\right)+
\left(\vec{{\bf A}}'\,\vec{ p}\,'\right)+
h\left(\vec{{\bf H}}\,'\,\vec{{\sigma}}\,'\right)\right]+
$$
$$
+\frac{e^2 hc(-i)}{{(E+E_0 +e\varphi)}^2}
\left(\vec{\sigma}\,'\,\vec{{\bf A}}'\right)
\left(\vec{{\sigma}}\,'\,\vec{ {\bf E}}\,'\right)=
-\frac{e}{cI}\left[
\frac{\left(\vec{J}\,\vec{\mu}_1\right)(\vec{r}_1\,\vec{r})}{r_1^3}-
\frac{\left(\vec{J}\,\vec{r}_1\right)(\vec{\mu}_1\,\vec{r})}{r_1^3}+
\right.
$$
$$
\left.
+\frac{\left(\vec{J}\,\vec{\mu}_2\right)(\vec{r}_2\,\vec{r})}{r_2^3}-
\frac{\left(\vec{J}\,\vec{\mu}_2\right)(\vec{\mu}_2\,\vec{r})}{r_2^3}+
\right]+
\frac{ec}{E+E_0 +e\varphi}\left[
h\left(\vec{{\bf H}}\,\vec{{\sigma}}\,'\right)
+\left(\vec{p}\,'\,\vec{{\bf A}}\right)+
\left(\vec{{\bf A}}\,\vec{ p}\,'\right)+
\right.
$$
$$
\left.
+\frac{1}{cI}\left(
\vec{J}\,\left[\vec{p}\,'\,\vec{r}\,\right]\right)\varphi +
\frac{h}{cI}\left\{
\left(\vec{J}\,\vec{{\bf E}}\right)\left(\vec{\sigma}\,'\,\vec{r}\,\right)-
\left(\vec{J}\,\vec{\sigma}\,'\,\right)\left(\vec{{\bf E}}\,'\,\vec{r}
\right)\right\}\right]+
$$
$$
+\frac{e^2 hc(-i)}{{(E+E_0 +e\varphi)}^2}
\left[\left(\vec{\sigma}\,'\,\vec{{\bf E}}\right)
\left(\vec{{\sigma}}\,'\,\vec{ {\bf A}}\right)+
\frac{\varphi}{cI}
\left(\vec{\sigma}\,'\,\vec{{\bf E}}\right)\left(\vec{\sigma}\,'\,\left[
\vec{J}\,\vec{r}\,\right]\right)\right.+
$$
$$
+\left.\frac{\left( \vec{\sigma}\,'\,\vec{{\bf A}}\right)}{cI}\left\{
\left(\vec{\sigma}\,'\,\vec{J}\,\right)\left(\vec{{\bf H}}\,\vec{r}\,\right)-
\left(\vec{{\bf H}}\,\vec{J}\,\right)\left(\vec{\sigma}\,'\,\vec{r}\,
\right)\right\}\right].
$$
The ortho--para transitions prohibition breaking is provided by
the odd--odd part of operator ${(V_{e-n})}_{magn}$. One can see from
${(V_{e-n})}_{magn}$, that obtained by taking into account the rotations
additional to (6) terms, which gives the contribution to $W'$,
are inversely proportional to $m_p^2$.

Therefore, provided by rotations the electron--nuclei interaction,
which causes the ortho--para transitions,
is proportional to $m_p^{-2}$.
 It is clear, that the direct magnetic interaction of protons
 has the same behaviour.
Thus, one can think, that the rotations neglecting
in interaction operator $V_{aa}$ gives a mistake of the same
order as $V_{n-n}$ neglecting.

At last, spin--angular interaction of electrons is contained in operator
$V_{ss}$ and doesn't influence on the possibility of ortho--para
transitions. It being known, that the contribution of spin--angular
interaction to $V_{ss}$ is insignificant in comparison with one
provided by rotation. The reason is, that in light molecules, as usual,
we have Hund's b-type connection. In this case, the influence
of rotation on electron motion is stronger that the spin--angular
interaction.

In the case of molecular ion of hydrogen one can estimate
the weight coefficients if we add the wave functions of opposite
symmetry to the unperturbed wave functions of ortho-- and para--states.

The correction to the wave function of unperturbed para--state
$\psi_{000}^-$  due to magnetic interaction between electron
and nuclei (8) in the second order of perturbation theory has the form
$$
\varphi^{+}=\sum\limits_{pvr}
\frac{<\psi_{pvr}^{+}|W_{aa}|\psi_{000}^{-}>}{E_{000}^- -E_{pvr}^+}
\psi_{pvr}^+ ,
$$
$$
W_{aa}=\frac{1}{4}\left[W(R_1 R_2 \Sigma_1 \Sigma_2)
-W(R_2 R_1 \Sigma_1 \Sigma_2)-W(R_1 R_2 \Sigma_2 \Sigma_1)+
W(R_2 R_1 \Sigma_2 \Sigma_1)\right].
$$
In the matrix elements of the sum the production of coordinate
wave functions of electron contains the  coordinates of nuclei
as a parameters. At that, in diagonal on electron state matrix element
the production of electron wave functions is even function
relative on permutations of spatial coordinates of nuclei,
whereas the operator $W_{aa}$ is odd function on definition.
Therefore, the matrix element diagonal on ground electron state
is equal to zero in spite of uncompensated spin of electron in
molecular ion and not equal to zero average quantity of the interaction
operator $W$ between nuclei spins and spin of electron on the ground state.
Thus, the main contribution to the wave function correction has the
form
$$
\varphi^{+}\sim
\frac{<\psi_{101}^{+}|W_{aa}|\psi_{000}^{-}>}{\Delta E}\psi_{101}^{+}.
$$
At that, $\Delta E$ is the difference in energy between electron states,
and $\Delta E =10eV$. The perturbation operator (8) has an order of
hyperfine structure in hydrogen atom and corresponds to the frequency
of 1400MHz. Therefore,
$$
<\psi_{001}^{+}|W_{aa}|\psi_{000}^{-}>\sim h1400\approx 10^{-6}eV,\quad
\varphi^{+}\sim 10^{-7}\psi_{101}^{+}.
$$
Thus, for the correction to the ortho--state wave function  we have
$$
\varphi^- \sim 10^{-7}\psi_{100}^{-}.
$$
The direct dipole--dipole interaction of nuclei $V_{n-n}$
is proportional to $m_{p}^{-2}$ and the interaction
constant for hydrogen molecule is approximately equal to 57 kHz.
Then, the direct nuclear interaction in hydrogen molecular ion
should be the same on value, that is much smaller, than the hyperfine
interaction constant between electron and nuclei. Thus, on calculating
the ortho-- and para--state wave function corrections we can
neglect the interaction $V_{n-n}$ in comparison with $V_{e-n}$ in
the expression for $V_{aa}$.

So, one can calculate the perturbed wave functions for spontaneous
ortho--para transitions in ion $H_2^{+}$ dealing with electron--nuclei
interaction operator $W_{aa}$, which is proportional to $m_p^{-1}$
and antisymmetrical relative to the permutations of spatial
or spin coordinates.
At that, we don't take into account an interactions bilinear
on nuclear spin and a distortion of wave functions in zero
approximation order due to the operator $V_{ss}$, which doesn't provide
the ortho--para transitions prohibition breaking.
\begin{center}
{\bf\large Summary}
\end{center}
Spontaneous nuclear ortho--para transitions are shown to be possible
in hydrogen molecule and molecular ion as due to hyperfine interaction
odd-odd relative to the space or spin nuclear coordinate permutations.

A part of this interaction inversely proportional to the first power
of nuclear mass is found for hydrogen molecular ion.
\begin{center}
{\bf \large Acknowledgments}
\end{center}
I would like to thank Prof. P. P. Pavinsky for the subject of the
work and help.\\
\begin{center}
{\bf \large References}\\
\end{center}

[1] K. F. Bonhoeffer, P. Harteck, Zs. f. Phys. Chem., {\bf B4}, 113, 1929.\\

[2] A. Farcas, Ortho-- para-- and heavy hydrogen., Moscow, ONTI, 1936.\\

[3] K. Motizuki, T. Nagamiya, J. Phys. Soc. Japan, {\bf 11}, N2, 93, 1956.\\

[4] H. Bethe, Quantum mechanics of simplest systems, Moscow-Leningrad,

ONTI, 1935.

\end{document}